\documentclass[conference,a4paper]{IEEEtran}
\IEEEoverridecommandlockouts

\usepackage{cite}
\usepackage{amsmath,amssymb,amsfonts}
\usepackage{url}

\usepackage{xcolor}
\usepackage{balance}

\ifCLASSINFOpdf
  \usepackage[pdftex]{graphicx}
\else
  \usepackage[dvips]{graphicx}
\fi
\ifCLASSOPTIONcompsoc
 \usepackage[caption=false,font=normalsize,labelfont=sf,textfont=sf]{subfig}
\else
 \usepackage[caption=false,font=footnotesize]{subfig}
\fi
\begin{document}
%
\title{Power-Efficient Beam Pattern Synthesis via Dual Polarization Beamforming}

\author{\IEEEauthorblockN{Sven O. Petersson} 
\IEEEauthorblockA{  
\textit{Ericsson Research, Gothenburg, Sweden, sven.petersson@ericsson.com}
}                                     

}



\maketitle

\begin{abstract}
Beamforming is traditionally associated with coherent summation of signals from antenna elements of the same polarization, here referred to as single polarization beamforming (SPBF). In this paper we focus on a new method, called dual polarization beamforming (DPBF), to design beam patterns. Instead of using only a single element-polarization, as for SPBF,  a desired beam pattern is designed as the sum of powers for two orthogonal element-polarizations. So, with DPBF the focus is on \textsl{total} radiated power beam patterns. The DPBF technique provides additional degrees of freedom to form a desired beam pattern such that amplitude variations in the beamforming vector can often be significantly reduced, potentially to uniform amplitude. Phase-only beamforming, possibly using only minor amplitude variations, is a very interesting property, especially for active antennas, since it offers the potential of full power amplifier utilization. In this paper we apply DPBF to uniform linear arrays (ULAs) as well as uniform rectangular arrays (URAs). In communication systems, it is often desired to have a second beam with identical beam pattern but orthogonal polarization in all directions. We show how such beams are  designed with DPBF, both for ULAs and URAs.
\end{abstract}

\vskip0.5\baselineskip

%

\section{Introduction}
\label{sec:introduction}
A majority of antennas used in cellular systems are two-port antennas, the ports representing beams with identical power patterns and orthogonal polarization. To improve spectral efficiency, two-port antennas are increasingly replaced with multi-port antenna panels. This allows for user-specific beamforming with narrow beams, or more generally beams adapted to the channel between the base station (BS) and the user equipment (UE).
However, not all channels in cellular systems are to be beam formed to a specific user \cite{3GPP:36.211}. Some channels like cell-specific reference signal (CRS), Physical Broadcast Channel (PBCH), Primary Synchronization Signal (PSS) and Secondary Signal Channel (SSS) are non-user specific and are to be transmitted over a wider area.
It is well known that the smallest halfpower beamwidth, $\phi_{3dB}$, that can be generated by an antenna is inversely proportional to the array size in wavelengths, $D/\lambda$. For uniform amplitude and phase tapers, and array sizes of 1$\lambda$ or larger, this halfpower beamwidth is approximately given by
\begin{equation}\phi_{3dB}=0.88/(D/\lambda).\label{equ:hpbw}\end{equation}
%
%
By introducing non-uniform amplitude and/or  phase taper one can create wider beams than what is given by \eqref{equ:hpbw}. For passive arrays amplitude taper is a matter of rerouting power, whereas for active antennas amplitude taper is undesired since it will reduce output power. An example of amplitude taper can be found in \cite[Table 1]{Hamdy2017} where the weight for a broadcast beam of 65$^\circ$ half power beamwidth suffers from a 2.0 dB reduction in output power due to amplitude taper. Another example is \cite[Table 1]{Foo2008} where a broadcast beam of 65$^\circ$ half power beamwidth suffers from a 3.0 dB reduction in output power and yet another example is \cite{Qiao2016}. The importance of power efficient beamforming is well known \cite{Fu2013}. A straightforward method to achieve good power utilization is to apply phase taper only, i.e., equal power per antenna. However, phase taper only often results in the resulting beam pattern having significant ripple \cite{Fu2013}, \cite{Yuan2017}. In addition, for a two antenna case phase taper can not be used at all since that will change the beampointing direction.

This paper presents a power-efficient beam pattern synthesis technique, dual-polarization beamforming (DPBF), which offers additional degrees of freedom in the beamforming process. By this it is possible to generate beams significantly wider than what is shown by \eqref{equ:hpbw}, in many cases, by means of phase taper only, thus enabling efficient use of power resources. The technique is perhaps most important for sector covering, or any other wide, beams which already suffer from lower antenna gain compared to user specific beams because of the difference in beamwidths. User specific beams, for example codebook based transmission in \cite{3GPP:36.213}, are in many cases based on linear phase and uniform amplitude taper resulting in good power utilization.

The concept of DPBF was first described in a patent application, \cite{P30312_IntPubl}, published in 2011. The same technique applied to a sparse array was presented in another patent application, \cite{P29518_IntPubl}, also published in 2011. In \cite{Tzanidis2015}, some of the ideas described in \cite{P29518_IntPubl} are applied to a 4x8 sparse array. There are however no explanation in \cite{Tzanidis2015} regarding the methodology used when designing the single beam pattern but an analysis of the presented phase taper, for the single beam, reveals that DPBF is applied in the horizontal domain (8 elements) whereas classical beamforming is applied in elevation. 

The rest of the paper is organized as follows. In section \ref{sec:antennas}, examples of antenna configurations suitable for DPBF are presented. In section \ref{sec:beamforming for ULA}, DPBF is applied to a ULA and in section \ref{sec:beamforming for URA}, to a URA, followed by conclusions in section \ref{sec:conclusion}.  

\section{Antennas}
\label{sec:antennas}
\begin{figure}[t!]
\centering
\includegraphics[width = .9\columnwidth]{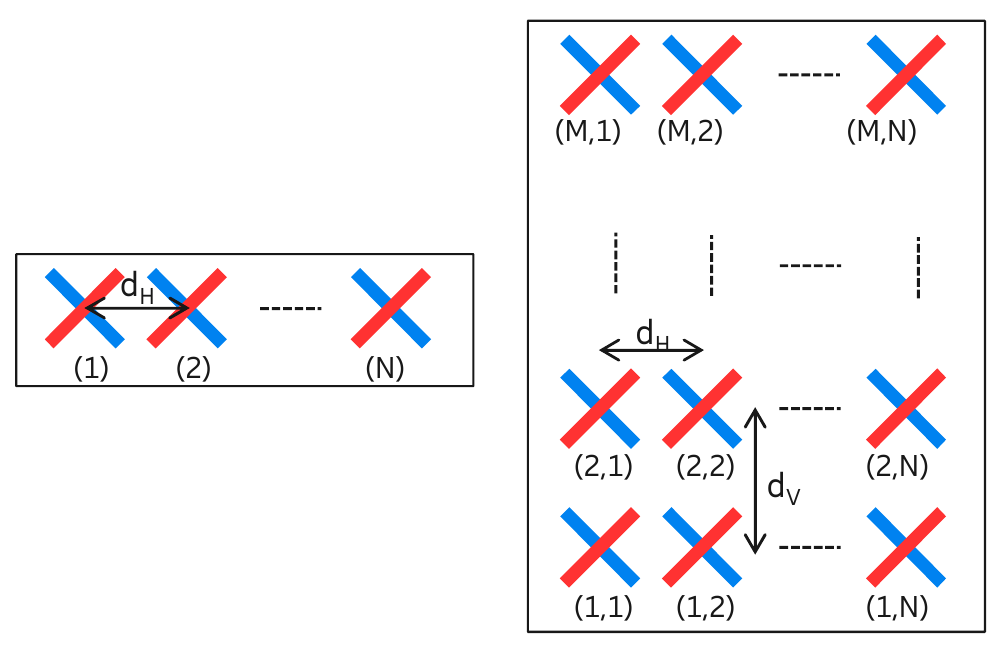}
\caption{Dual polarized arrays. $N$-element ULA (left) and $M\times N$-element URA (right).}
\label{fig:Arrays}
\end{figure}

The scope of this paper is limited to ideal dual polarized antenna arrays, specifically uniform linear arrays (ULAs) and uniform rectangular arrays (URAs). Such arrays are populated by multiple, identical dual polarized elements located at, per dimension, equal distances. Polarizations are assumed orthogonal and power patterns for the two polarizations are assumed identical. Although the description herein assumes dual polarized elements, the technique is equally suitable also for two arrays with identical topology where the different arrays contain elements with orthogonal polarizations. Fig.~\ref{fig:Arrays} show two examples of arrays in which the blue and red elements symbolically indicate element positions as well as element polarizations A (red) and B (blue). The steering vector for a ULA can be expressed as
\begin{equation}
\mathbf{a}(\theta,\phi)=[     {a}_{1}(\theta,\phi),   \cdots   , {a}_{N}(\theta,\phi) ]^T,\label{eq:steering_vec} 
\end{equation}
where $\mathbf[\cdot]^T$ denotes the transpose of a matrix and ${a}_{n}(\theta,\phi)$ represents phase shift of the signal at the $n$th antenna relative, for example, the center of the array in the direction given by ($\theta$,$\phi$), as well as amplitude.

\section{Beamforming for ULA}
\label{sec:beamforming for ULA}
This section describes classical beamforming, herein named single polarization beamforming (SPBF), as well as the novel concept of DPBF. The techniques are described by means of a four-column ULA example. It is shown that, given the beamforming weights for a first beam, the beamforming weights for a second beam can be designed such that the two beams have identical power patterns and orthogonal polarizations in all directions.

\subsection{Single Polarization Beamforming}
\begin{figure}[t!]
\centering
\includegraphics[width = .9\columnwidth]{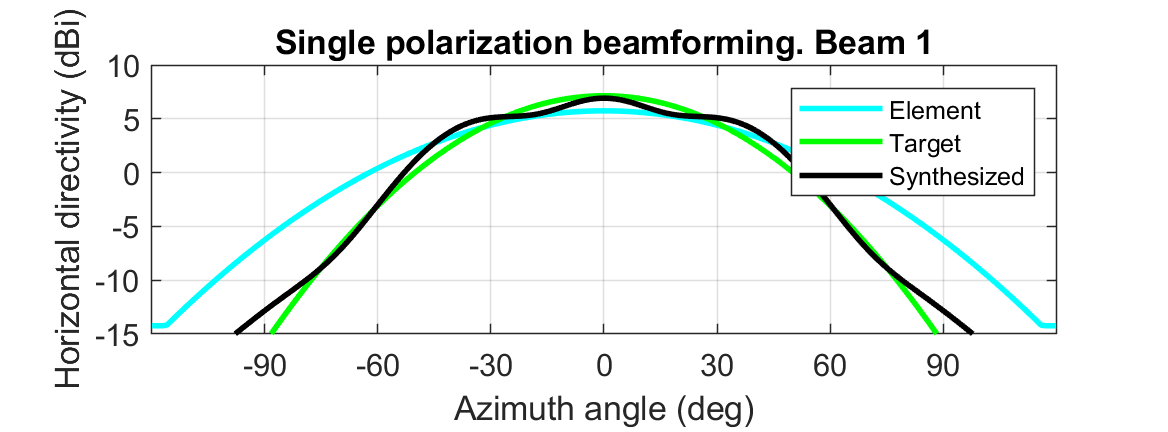}
\caption{Sector pattern created by means of SPBF.}
\label{fig:SPBF_Sectorpatt}
\end{figure}

An example of a synthesized beam pattern formed by an ideal ULA with parameters according to Table \ref{table:ArrayParam} is shown in Fig.~\ref{fig:SPBF_Sectorpatt}. The beam pattern has been found by means of multi-objective optimization with variance for the difference in dB between the target and the synthesized patterns as a first cost and amplitude taper, representing output power loss, as a second. 

\begin{table}
\caption{Array parameters}
\label{table:ArrayParam}
\setlength{\tabcolsep}{3pt}
\begin{tabular}{|p{150pt}|p{70pt}|}
\hline
Number of columns & $N=4$  \\
\hline
Number of rows (for URA) & $M=6$  \\
\hline
Element pattern shape (both polarizations) & Gaussian \\
\hline
Polarizations & Orthogonal \\
\hline
Element half-power beamwidth & 90$^\circ$ \\
\hline
Column separation $\textnormal{d}_\textnormal{H} $& 0.5 lambda \\
\hline
Row separation  $\textnormal{d}_\textnormal{V}$ & 0.7 lambda \\
\hline
Target beam pattern shape & Gaussian \\
\hline
Target half-power beamwidth & 65$^\circ$  \\
\hline
\end{tabular}
\label{tab1}
\end{table}
The beamforming vector for a first beam, feeding all elements of polarization A (red in Fig.~\ref{fig:Arrays}), resulting in the synthesized beam pattern shown in Fig.~\ref{fig:SPBF_Sectorpatt}  is
\begin{equation}\mathbf{w}_{1,\textnormal{A}}=[ 1.0,1.0,-0.48,0.24]^T.\label{eq:w1_SPBF} \end{equation}
Clearly, this vector is just one example but anyway serves as a reasonable choice in trading pattern shape and power utilization. As can be seen from \eqref{eq:w1_SPBF} there is a significant amount of amplitude taper corresponding to an output power drop, or weighting loss, of 2.4 dB compared to a case where the magnitude for all weights equals 1. It shall be noted that all patterns in Fig.~\ref{fig:SPBF_Sectorpatt} are  normalized such that the total radiated power for the azimuth cut equals 2$\pi$ for each of the three patterns. Thus, the actual transmit power is not reflected, only the shape of the patterns.

\subsection{Dual Polarization Beamforming}

\begin{figure}[t!]
\centering
\includegraphics[width = .9\columnwidth]{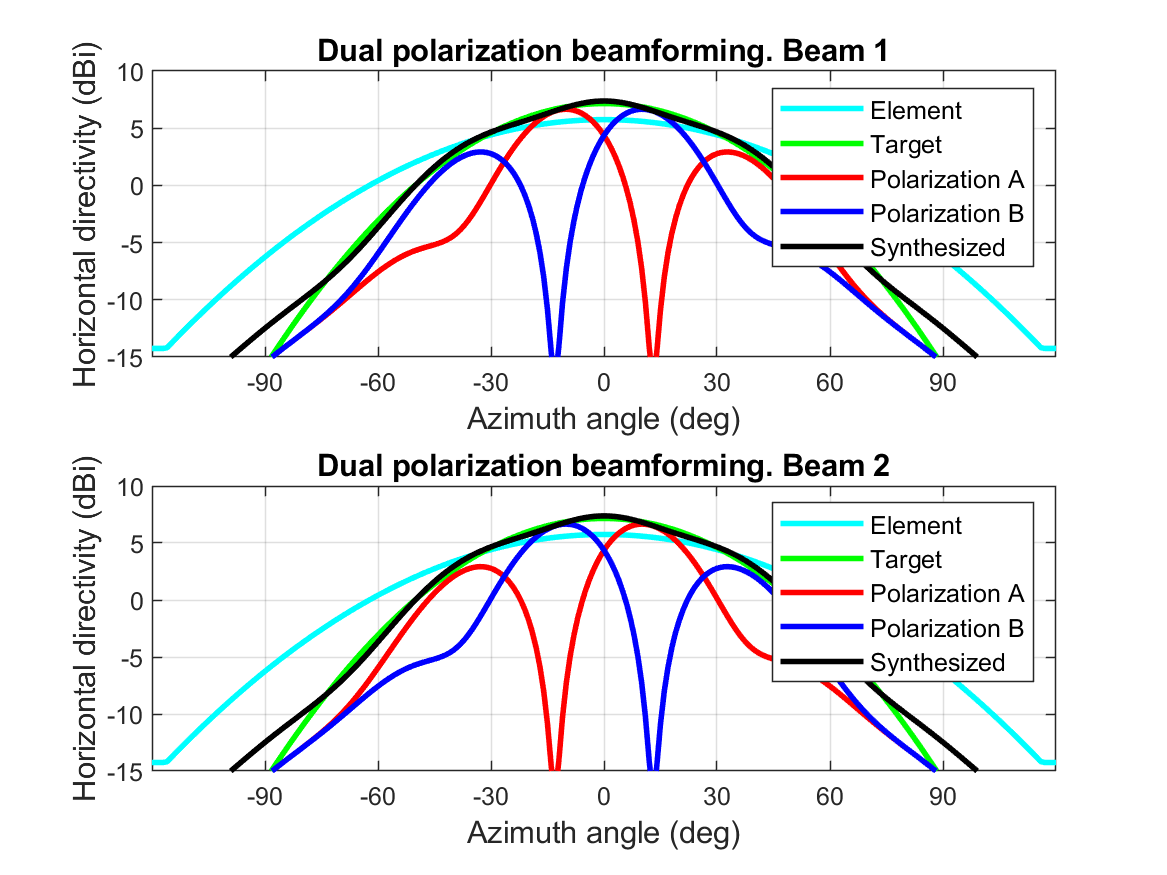}
\caption{Sector patterns created by means of DPBF.}
\label{fig:DPBF_Sectorpatt}
\end{figure}

The basic idea of DPBF is to design a beam pattern such that the total power pattern, i.e., the sum of the power patterns per polarization, gives the desired beam shape. Fig.~\ref{fig:DPBF_Sectorpatt}, top, shows an example of a pattern for a first beam created via DPBF for the same array as for the SPBF example, again found via multi-objective optimization. For simplicity it was assumed that the beamforming weight for the B-polarization is the complex conjugate of the beamforming weight for the A-polarization. The Element, Target and Synthesized patterns are all directivity normalized to have total radiated power equal to 2$\pi$, whereas the Polarization A and Polarization B patterns are normalized to have total radiated power equal to $\pi$ each. Thus, the patterns do not reflect the significant gain in transmitted power compared to SPBF. 
The resulting beam patterns, in black, in Fig.~\ref{fig:SPBF_Sectorpatt} and Fig~\ref{fig:DPBF_Sectorpatt}, top, are quite similar but the pattern created via DPBF show a better correspondence (lower first cost) with the desired beamshape. And, more important, the DPBF pattern is created by means of phase taper only so the weight loss equals 0 dB. The beamforming weights, one weight each for the orthogonal polarizations A and B, are for a first beam given as
\begin{equation}
\mathbf{w}_{1,\textnormal{A}}=[ e^{i2.32}, e^{i2.06},e^{i0.00},e^{i0.97}   ]^T\label{equ:w1A_DPBF} 
\end{equation}
\begin{equation}
\mathbf{w}_{1,\textnormal{B}}=\mathbf{w}_{1,\textnormal{A}}^*=[ e^{-i2.32}, e^{-i2.06},e^{-i0.00},e^{-i0.97}   ]^T\label{qu:w1B_DPBF} 
\end{equation}
where $\mathbf(\cdot)^*$ is the complex conjugate.
The DPBF technique is not limitated to the use of phase taper only. In some cases, amplitude taper might be needed as well to add degrees of freedom but then the amplitude taper is typically significantly less than for SPBF.
As can be seen from Fig.~\ref{fig:DPBF_Sectorpatt} the polarization changes with direction. In some directions it may be identical to the inherent element polarization, at approximately $-15^\circ$  and $+15^\circ$, whereas in general it will be a weighted combination of the two intrinsic polarizations. As a consequence one cannot associate a beam generated via DPBF with a specific polarization state, something which is normally done for SPBF generated beams.

\subsection{Orthogonally Polarized Beam for ULA}
 
Despite the varying polarization for DPBF it is possible to create a second beam which has the same power pattern and, in each direction, orthogonal polarization relative the first beam.

For a ULA, with elements numbered according to Fig.~\ref{fig:Arrays}, i.e., from one side of the array to the other, the beamforming vectors for the second beam can be expressed in a very compact way as 
\vspace{-0.1cm}
\begin{equation}
\mathbf{w}_{2,\textnormal{A}}=-\mathbf{J}\mathbf{w}_{1,\textnormal{B}}^*
\label{eq:w2A}   
\end{equation}
\vspace{-0.4cm}
\begin{equation}
\mathbf{w}_{2,\textnormal{B}}=\mathbf{J}\mathbf{w}_{1,\textnormal{A}}^*
\label{eq:w2B}   
\end{equation}
where $\mathbf{J}$ is the exchange matrix, i.e., a permutation matrix which reverses the element order in the beamforming vectors for which
\begin{equation}
\mathbf{J}\mathbf{J}^H=\mathbf{I},
\label{eq:JJH}   
\end{equation}
where $\mathbf{I}$ is the identity matrix and $\mathbf(\cdot)^H$ is the hermitian transpose. 

The electrical field in direction ($\theta$,$\phi$), $\mathbf{e}_1$($\theta$,$\phi$), for a first beam corresponding to a first set of beamforming weights, $\mathbf{w}_{1,\textnormal{A}}$  and $\mathbf{w}_{1,\textnormal{B}}$, is given as

\vspace{-0.2cm}
\begin{equation}
\mathbf{e}_1(\theta,\phi)=
\begin{bmatrix}
    {e}_{1,\textnormal{A}}(\theta,\phi)  \\
    {e}_{1,\textnormal{B}}(\theta,\phi)
\end{bmatrix} =
\begin{bmatrix}
   \mathbf{w}_{1,\textnormal{A}}^T\mathbf{a}(\theta,\phi)  \\
    \mathbf{w}_{1,\textnormal{B}}^T\mathbf{a}(\theta,\phi)
\end{bmatrix},
\label{eq:E1}   
\end{equation}
where $\mathbf{a}$($\theta$,$\phi$)  is the steering vector for the array. Note that we have ignored the element pattern here, justified by the assumption of a ULA where all elements have the same power pattern.
Similarly, using \eqref{eq:w2A} and \eqref{eq:w2B}, the electrical field for the second beam in the same direction as the first beam, can be written as
\vspace{-0.2cm}
\begin{equation}
\mathbf{e}_2(\theta,\phi)=
\begin{bmatrix}
    \mathbf{w}_{2,\textnormal{A}}^T\mathbf{a}(\theta,\phi)  \\
    \mathbf{w}_{2,\textnormal{B}}^T\mathbf{a}(\theta,\phi)
\end{bmatrix} =
\begin{bmatrix}
    -\mathbf{w}_{1,\textnormal{B}}^H\mathbf{J}\mathbf{a}(\theta,\phi)  \\
    \mathbf{w}_{1,\textnormal{A}}^H\mathbf{J}\mathbf{a}(\theta,\phi)
\end{bmatrix}.
\label{eq:E2}   
\end{equation}
Polarization parallellity $\xi$ is defined as the magnitude of the inner product of the two fields 
\begin{equation}
\label{eq:pol_ort}
\xi = \left|\mathbf{e}_{1}^H(\theta,\phi)  \mathbf{e}_{2}(\theta,\phi) \right|
\end{equation}
and shall ideally be zero for all directions ($\theta$,$\phi$).
The inner product for the fields from \eqref{eq:E1} and \eqref{eq:E2} are found as  
\vspace{-0.1cm}
\begin{align}
\label{eq:Eort}
\mathbf{e}_{1}^H(\theta,\phi)  \mathbf{e}_{2}(\theta,\phi) =
&-\mathbf{a}^H(\theta,\phi)\mathbf{w}_{1,\textnormal{A}}^*\mathbf{w}_{1,\textnormal{B}}^H\mathbf{J}\mathbf{a}(\theta,\phi)\nonumber\\
&+\mathbf{a}^H(\theta,\phi)\mathbf{w}_{1,\textnormal{B}}^*\mathbf{w}_{1,\textnormal{A}}^H\mathbf{J}\mathbf{a}(\theta,\phi).
\end{align}
For a ULA, with element numbering according to Table \ref{table:ArrayParam}, we can use the equality 
\vspace{-0.3cm}
\begin{equation}
\label{eq:a_relation}
\mathbf{a}^*(\theta,\phi)=\mathbf{J}\mathbf{a}(\theta,\phi)
\end{equation}
in which we have assumed that the phase center is located at the center of the array.
Combining \eqref{eq:Eort} and \eqref{eq:a_relation} results in
\vspace{-0.3cm}
%

\begin{align}
\label{eq:Eort2}
\mathbf{e}_{1}^H(\theta,\phi)  \mathbf{e}_{2}(\theta,\phi)=
&-\mathbf{a}^H(\theta,\phi)\mathbf{w}_{1,\textnormal{A}}^*\mathbf{w}_{1,\textnormal{B}}^H\mathbf{Ja}(\theta,\phi)\nonumber\\
&+(\mathbf{a}^H(\theta,\phi) \mathbf{w}_{1,\textnormal{A}}^* \mathbf{w}_{1,\textnormal{B}}^H  \mathbf{Ja}(\theta,\phi))^T.
\end{align}

As the terms on the right-hand-side of \eqref{eq:Eort2} are scalars the sum  is always identical to zero.
This means that the polarization parallellity for the two fields is zero and the polarizations orthogonal to each other in all spatial directions.

In addition to orthogonal polarizations we also want the total power patterns, i.e., the sum of the partial (per polarization) power patterns, to be identical. The power in direction ($\theta$,$\phi$)  for the first beam is defined as
\begin{align}
\label{eq:Power1}
P_{1}(\theta,\phi)
&=\mathbf{e}_{1}^H(\theta,\phi)  \mathbf{e}_{1}(\theta,\phi)\nonumber\\
&=\mathbf{a}^H(\theta,\phi)(\mathbf{w}_{1,\textnormal{A}}^*\mathbf{w}_{1,\textnormal{A}}^T+\mathbf{w}_{1,\textnormal{B}}^*\mathbf{w}_{1,\textnormal{B}}^T)\mathbf{a}(\theta,\phi). \
\end{align}

Similarly the power pattern for the second beam is found from \eqref{eq:E2} and \eqref{eq:a_relation} to be

\vspace{-0.3cm}
\begin{align}
\label{eq:Power2}
{P}_{2}(\theta,\phi)
&=\mathbf{e}_{2}^H(\theta,\phi)  \mathbf{e}_{2}(\theta,\phi)\nonumber\\
&=(\mathbf{Ja}(\theta,\phi))^H(\mathbf{w}_{1,\textnormal{A}}\mathbf{w}_{1,\textnormal{A}}^H+\mathbf{w}_{1,\textnormal{B}}\mathbf{w}_{1,\textnormal{B}}^H)\mathbf{Ja}(\theta,\phi)\IEEEeqnarraynumspace\nonumber\\
&=\mathbf{a}(\theta,\phi)^T(\mathbf{w}_{1,\textnormal{A}}^T\mathbf{w}_{1,\textnormal{A}}^*+\mathbf{w}_{1,\textnormal{B}}^T\mathbf{w}_{1,\textnormal{B}}^*)^*\mathbf{a}(\theta,\phi)^*.
\end{align}

From \eqref{eq:Power1} and \eqref{eq:Power2}, and that power is real-valued, it follows that
\begin{equation}
\label{eq:P_relation}
{P}_{2}(\theta,\phi) = {P}_{1}(\theta,\phi)^*= {P}_{1}(\theta,\phi),\IEEEeqnarraynumspace
\end{equation}
thus, the power patterns for the two beams are identical in all directions ($\theta$,$\phi$).
The azimuth cut for a second beam corresponding to Fig.~\ref{fig:DPBF_Sectorpatt}, top,  is shown in Fig.~\ref{fig:DPBF_Sectorpatt}, bottom, The figure show that the partial power patterns are identical but that the polarization for the partial (per polarization) power patterns have changed.
%
%
\section{Beamforming for URA}
\label{sec:beamforming for URA}

A common technique to design a beamforming matrix for a URA is to design a weight vector for each of the two spatial array dimensions, assuming two separate, spatially orthogonal ULAs, and then combine these weight vectors to a two-dimensional (2D) weight matrix. This technique can be used for DPBF as well.
The beamforming matrices  $\mathbf{W}_{1,\textnormal{A}}$ and $\mathbf{W}_{1,\textnormal{B}}$, to be applied to the physical elements with polarization A and B respectively, are found as
\begin{equation}
\label{eq:W1A}
\mathbf{W}_{1,\textnormal{A}}=\mathbf{u}_{1,\textnormal{A}}\mathbf{v}_{1,\alpha}^T+\mathbf{u}_{2,\textnormal{A}}\mathbf{v}_{1,\beta}^T
\end{equation}
\begin{equation}
\label{eq:W1B}
\mathbf{W}_{1,\textnormal{B}}=\mathbf{u}_{1,\textnormal{B}}\mathbf{v}_{1,\alpha}^T+\mathbf{u}_{2,\textnormal{B}}\mathbf{v}_{1,\beta}^T
\end{equation}
where $\mathbf{u}_{1,\textnormal{A}}$ and $\mathbf{u}_{1,\textnormal{B}}$ represent the beamforming vectors, for a first beam, to be applied in elevation. These vectors can be seen as forming a new, virtual, element with polarization $\alpha$. Similarly $\mathbf{u}_{2,\textnormal{A}}$ and $\mathbf{u}_{2,\textnormal{B}}$, derived from $\mathbf{u}_{1,\textnormal{A}}$ and $\mathbf{u}_{1,\textnormal{B}}$, represent the elevation beamforming vectors, for a second and orthogonally polarized element with polarization $\beta$. The vectors  $\mathbf{v}_{1,\alpha}$ and $\mathbf{v}_{1,\beta}$ defines the beamforming in azimuth and is applied to the, virtual, elements with polarizations $\alpha$ and $\beta$ respectively.

The vectors $\mathbf{u}_{2,\textnormal{A}}$ and $\mathbf{u}_{2,\textnormal{B}}$ can be derived from $\mathbf{u}_{1,\textnormal{A}}$ and $\mathbf{u}_{1,\textnormal{B}}$ from \eqref{eq:w2A} and \eqref{eq:w2B} giving that  \eqref{eq:W1A} and \eqref{eq:W1B}  can be simplified to contain only beamforming vectors for a first beam, 
\vspace{-0.2cm}
\begin{equation}
\label{eq:W1A_2}
\mathbf{W}_{1,\textnormal{A}}=\mathbf{u}_{1,\textnormal{A}}\mathbf{v}_{1,\alpha}^T-\mathbf{Ju}_{1,\textnormal{B}}^*\mathbf{v}_{1,\beta}^T
\end{equation}
\begin{equation}
\label{eq:W1B_2}
\mathbf{W}_{1,\textnormal{B}}=\mathbf{u}_{1,\textnormal{B}}\mathbf{v}_{1,\alpha}^T+\mathbf{Ju}_{1,\textnormal{A}}^*\mathbf{v}_{1,\beta}^T.
\end{equation}

\subsection{DPBF in both elevation and azimuth}
\begin{figure}[t!]
\centering
\includegraphics[width = .9\columnwidth]{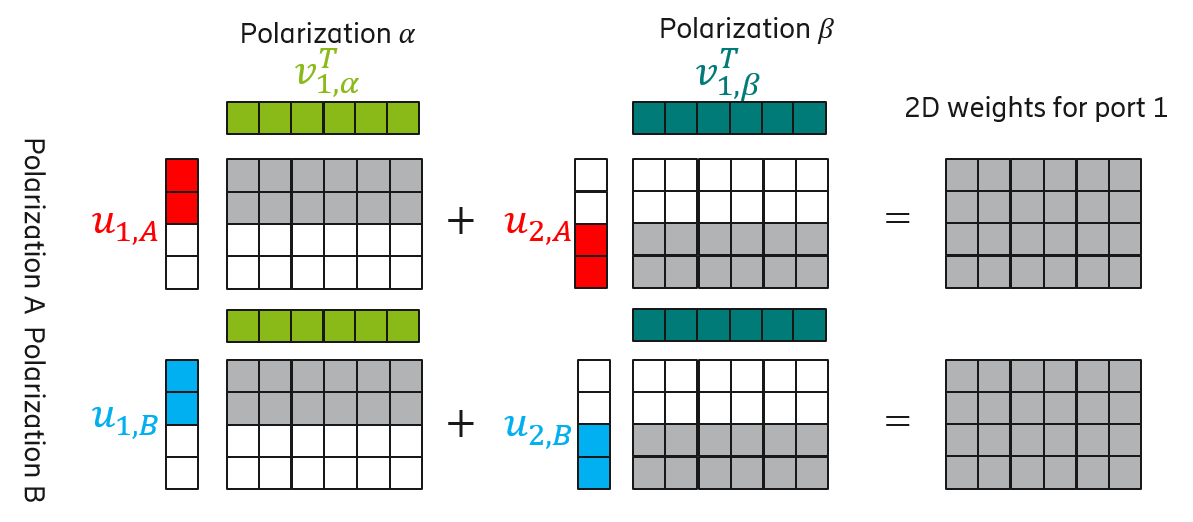}
\caption{Element usage for 2D array with DPBF along both directions.}
\label{fig:2D_ElementUsage}
\end{figure}

To avoid undesired coherent combining, which typically results in poor use of PA power or even worse can lead to overload of PAs, when combining the 1D vectors to 2D matrices the 1D vectors must be designed such that some elements have zero-power (ZP). One example of such design is shown in Fig. \ref{fig:2D_ElementUsage} which is representing equations \eqref{eq:W1A} and \eqref{eq:W1B}. In the example the $\mathbf{u}$-vectors for elevation beamforming are designed such that they, in both polarizations, have non-zero power (NZP) in the first $M$/2 vector elements, indicated by the red and blue colors, and zero power (ZP) in the remaining vector elements, indicated by the white color. In the $\mathbf{v}$-vectors, for azimuth beamforming, all $N$ vector elements have NZP, indicated by the light and dark green colors. As can be seen in the upper half of the figure, representing \eqref{eq:W1A}, all antenna elements on the right hand side of the equal sign are grey indicating that all antenna elements for polarization A have NZP, i.e., are active. The grey elements on the left hand side do not overlap which means that there is only one contribution to each of the antenna elements. Thus, there will be no, from a power amplifier perspective, undesired coherent combination of the two contributions. If there was an overlap of the NZP portion of the $\mathbf{u}$-vectors, the signal fed on the antenna port would appear on some of the array elements as a sum of two weighted replicas. This would result in a change of signal amplitude on these elements potentially leading to PAs being overloaded. The above discussion applies as well for the lower half of Fig.~\ref{fig:2D_ElementUsage}, representing \eqref{eq:W1B}.

In this example we let the $\mathbf{u}$-vectors, here corresponding to elevation beamforming, contain the ZP elements but one can as well let the $\mathbf{v}$-vector contain these instead. Which one is preferred depends on the size of the array and how many degrees of freedom that is required to achieve the desired beamshape along each dimension.
\subsection{SPBF in elevation and DPBF in azimuth}
This case is very similar to the previous but, as the beamforming in elevation is based on SPBF we can allow all elements in one polarization to be NZP. All elements for the other polarization are ZP. The azimuth dimension is identical to the previous case. 
\begin{figure}[t!]
\centering
\includegraphics[width = .9\columnwidth]{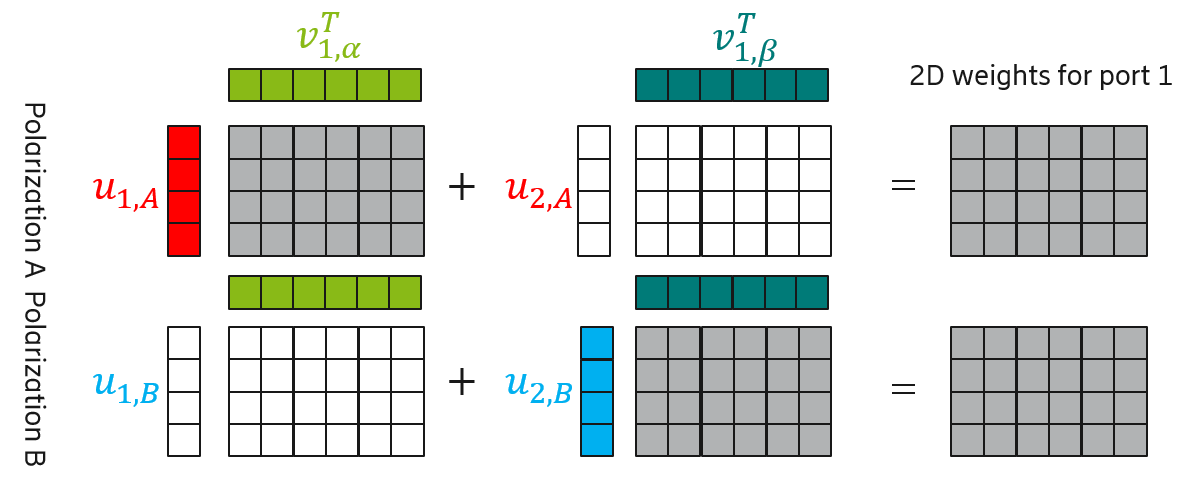}
\caption{Element usage for 2D array with SPBF in elevation and DPBF in azimuth.}
\label{fig:ElementUsageSPBF_DPBFt}
\end{figure}
Which dimension is selected for SPBF, the vertical or the horizontal, is a matter of array sizes, desired beamwidths in different dimensions, etc.

\subsection{Orthogonally polarized beam for URA}

For an URA, with elements numbered according to Fig.~\ref{fig:Arrays}, the beamforming vectors for the second beam is found in a very similar way as for the ULA case, the difference being that beamforming weights shall be reversed along two directions. For example the beamforming weight for beam 1 and polarization A applied to element ($m$,$n$) will for beam 2, after taking the complex conjugate, be applied to element ($M$-1-$m$, $N$-1-$n$) and polarization B. In matrix form this becomes
\begin{equation}
\label{eq:W2A}
\mathbf{W}_{2,\textnormal{A}}=-\mathbf{J}_M\mathbf{W}_{1,\textnormal{B}}^*\mathbf{J}_N
\end{equation}
\begin{equation}
\label{eq:W2B}
\mathbf{W}_{2,\textnormal{B}}=\mathbf{J}_M\mathbf{W}_{1,\textnormal{A}}^*\mathbf{J}_N
\end{equation}
where $\mathbf{J}_M$ is the exchange matrix of size $M\times M$, $\mathbf{J}_N$ is the exchange matrix of size $N\times N$.
To understand that \eqref{eq:W2A} and \eqref{eq:W2B} actually results in a second beam with identical power pattern and orthogonal polarization, compared to a first beam, one can start by looking at Fig.~\ref{fig:2D_ElementUsage} and apply the theory applicable for DPBF and ULA. The pair of vectors  $\mathbf{u}_{1,\textnormal{A}}$  and $\mathbf{u}_{1,\textnormal{B}}$ defines elements with polarization $\alpha$. Similarly, the pair of vectors  $\mathbf{u}_{2,\textnormal{A}}$ and $\mathbf{u}_{2,\textnormal{B}}$ defines elements that have the same power pattern but with orthogonal polarization $\beta$. Applying $\mathbf{v}_{1,\alpha}$ and $\mathbf{v}_{1,\beta}$ on the these elements results in a beam pattern for a first beam. By creating a new pair of vectors, $\mathbf{v}_{2,\alpha}$ and $\mathbf{v}_{2,\beta}$, based on $\mathbf{v}_{1,\alpha}$ and $\mathbf{v}_{1,\beta}$, via \eqref{eq:w2A} and \eqref{eq:w2B} and letting these new vectors operate on the elements with polarizations $\alpha$ and $\beta$,
\begin{equation}
\label{eq:W2A_2}
\mathbf{W}_{2,\textnormal{A}}=\mathbf{u}_{1,\textnormal{A}}\mathbf{v}_{2,\alpha}^T+\mathbf{u}_{2,\textnormal{A}}\mathbf{v}_{2,\beta}^T
\end{equation}
\vspace{-0.5cm}
\begin{equation}
\label{eq:W2B_2}
\mathbf{W}_{2,\textnormal{B}}=\mathbf{u}_{1,\textnormal{B}}\mathbf{v}_{2,\alpha}^T+\mathbf{u}_{2,\textnormal{B}}\mathbf{v}_{2,\beta}^T,
\end{equation}
 we get a second beam with the identical power pattern and orthogonal polarization. Again using \eqref{eq:w2A} and \eqref{eq:w2B} we get
\vspace{-0.1cm}
\begin{equation}
\mathbf{u}_{1,\textnormal{A}}=\mathbf{J}_M\mathbf{u}_{2,\textnormal{B}}^*
\end{equation} 
\vspace{-0.5cm}
\begin{equation}
\mathbf{u}_{2,\textnormal{A}}=-\mathbf{J}_M\mathbf{u}_{1,\textnormal{B}}^*
\end{equation}  
\vspace{-0.5cm}
\begin{equation}
\mathbf{v}_{2,\alpha}^T=-\mathbf{v}_{1,\beta}^H\mathbf{J}_N
\end{equation} 
\vspace{-0.5cm}
\begin{equation}
\mathbf{v}_{2,\beta}^T=\mathbf{v}_{1,\alpha}^H\mathbf{J}_N,
\end{equation}  
which combined with \eqref{eq:W1B} results in
\vspace{-0.2cm}
\begin{align}
\mathbf{W}_{2,\textnormal{A}}
&=-\mathbf{J}_M(\mathbf{u}_{2,\beta}^*\mathbf{v}_{1,\beta}^H+\mathbf{u}_{1,\beta}^*\mathbf{v}_{1,\alpha}^H)\mathbf{J}_N\nonumber\\
&=-\mathbf{J}_M\mathbf{W}_{1,\textnormal{B}}^*\mathbf{J}_N
\end{align}
In a similar way one can easily show that \eqref{eq:W2B} holds.

\begin{figure}[t!]
\centering
\includegraphics[width = .9\columnwidth]{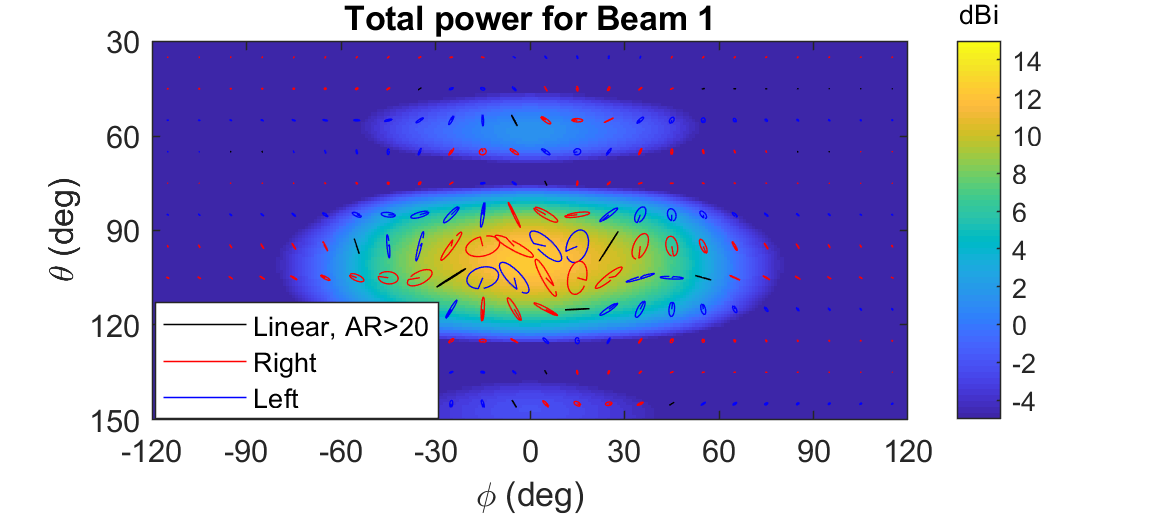}
\caption{Total power pattern for a first beam synthesized via DPBF in both elevation and azimuth.}
\label{fig:2D_Beam1}
\end{figure}

\begin{figure}[t!]
\centering
\includegraphics[width = .9\columnwidth]{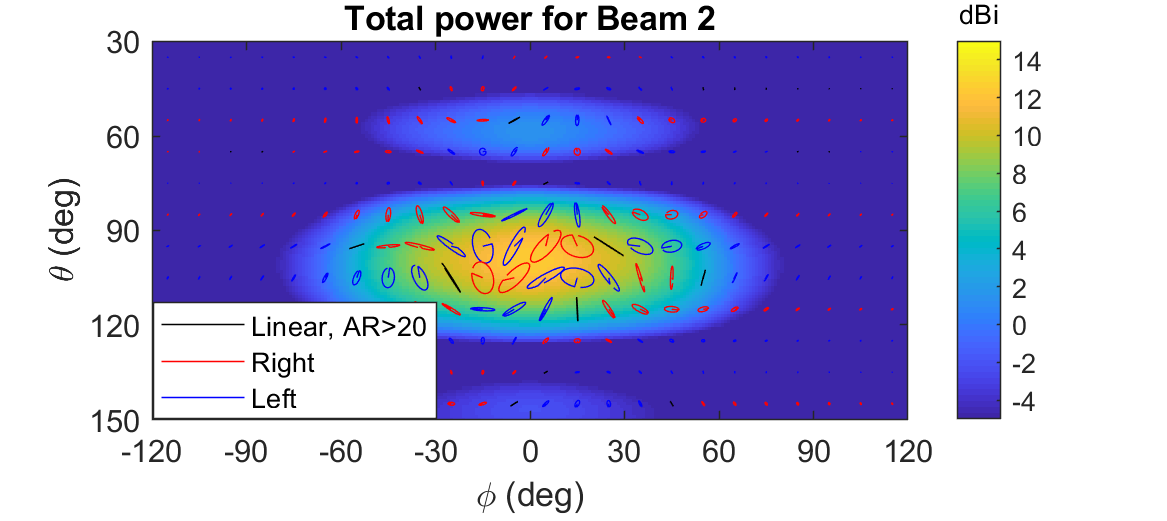}
\caption{Total power pattern for a second, orthogonally polarized, beam synthesized via DPBF in both elevation and azimuth.}
\label{fig:2D_Beam2}
\end{figure}

An example of 2D beamforming via an URA, applying DPBF in both elevation and azimuth, is shown in Fig.~\ref{fig:2D_Beam1} for beam 1 and Fig.~\ref{fig:2D_Beam2} for beam 2. The power patterns for the two beams are identical and the polarizations are orthogonal in all directions. To visualize the latter, polarization ellipses are overlaid the power patterns. When the axis ratio (AR in the legend) for a polarization ellipse exceeds 20 the polarization is shown as linear.

\section{Conclusion}
\label{sec:conclusion}
In this paper we have shown that beam patterns, with large beamwidths in relation to the array size, can be synthesized by means of phase taper only using the DPBF technique. Phase taper only is a very important aspect for active antennas since it offers efficient use of available power resources, in contrast to amplitude taper. Thus DPBF is a power-efficient beam pattern synthesis technique. We have also shown that, despite the fact that the polarization for such a beam changes with spatial angles, it is possible to design pairs of beams where the power patterns are identical and polarizations orthogonal, in all directions of interest.

\section{Acknowledgment}
\label{sec:Acknowledgment}
The author wants to thank Stefan F. Johansson and Martin N. Johansson for fruitful discussions, specifically on issues related to polarization, which inspired this work and also other colleauges that have helped improving this paper.

\bibliographystyle{IEEEtran}
\bibliography{storsven}

\end{document}